# Thermoelectric probe of defect state induced by ionic liquid gating in vanadium dioxide

Hui Xing[1, 2 *], Peihong Zhang[2], and Hao Zeng[2, †]

[1]*Key Laboratory of Artificial Structures and Quantum Control and Shanghai Center for Complex Physics, School of Physics and Astronomy, Shanghai Jiao Tong University, Shanghai 200240, China*

[2]*Department of Physics, University at Buffalo, SUNY, Buffalo, New York 14260, United States*

**Thermoelectric measurements detect the asymmetry between the density of states above and below the chemical potential in a material. It provides insights into small variations in the density of states near the chemical potential, complementing electron transport measurements. Here, combined resistance and thermoelectric power measurements are performed on vanadium dioxide ($VO_2$), a prototypical correlated electron material, under ionic-liquid (IL) gating. With IL gating, charge transport below the metal-to-insulator-transition (MIT) temperature remains in the thermally activated regime, while the Seebeck coefficient exhibits an apparent transition from semiconducting to metallic behavior. The contrasting behavior indicates changes in electronic structure upon IL gating, due to the formation of oxygen defect states. The experimental results are corroborated by numerical simulations based on a model density of states incorporating a gating induced defect band. This study reveals thermoelectric measurements to be a convenient and sensitive probe for the role of defect states induced by IL gating in suppressing the MIT in $VO_2$, which remains benign in charge transport measurements, and possibly for studying defect sates in other materials.**

\* huixing@sjtu.edu.cn
† haozeng@buffalo.edu

A metal-insulator transition (MIT) coupled with a structural transition at around 340 K [1] in vanadium dioxide ($VO_2$), a the prototypical strongly correlated system, has drawn extensive attention both in the study of correlated physics [2] and in practical applications such as field-effect transistors and ultrafast optoelectronic switches [3]. While the detailed physics behind the MIT [2, 4] is still controversial, it is generally recognized that the dimerization of vanadium ions is responsible for the structural transition and the accompanying change in electronic structures [5]. With vanadium dimerization, $VO_2$ transforms from a high temperature metallic rutile phase (R phase) to a low-temperature semiconducting monoclinic phase (M1 phase), which opens a gap of ~ 0.6 eV [6]. The driving force and the dynamics underlying the MIT in this correlated system have been under intensive investigation [7]. Thermal transport studies on pristine $VO_2$ have revealed intriguing behaviors in both metallic state [8] and the MIT transition regime [9]. Significant efforts in modulating the MIT for electronic applications were also pursued [10].

Recently, ionic-liquid (IL) gating has been applied in a wide range of materials and devices [10b, 11]. Under a gate bias, an electric double layer (EDL) is formed [12] leading to high electric fields capable of producing up to $10^{15}$ cm$^{-2}$ carriers [13]. Upon IL gating, MIT in $VO_2$ can be strongly modulated and even completely suppressed [14]. While this was previously believed to be an electrostatic gating effect, later experiments favor the formation of oxygen vacancies as the main mechanism responsible for the observed effects [10a, 15]. Unfortunately, the oxygen vacancy mechanism evaded detection in earlier studies, as conventional charge transport studies cannot distinguish electrostatic gating effect from gating induced change in defect states. A more convenient and direct probe of such defect states due to IL gating is needed.

Here we report results on combined thermoelectric and charge transport measurements of IL-gated $VO_2$. Utilizing its sensitivity to the asymmetry of the density of states (DOS) near the chemical potential, the measurement of thermoelectric power (Seebeck coefficient) combined with a simultaneous measurement of resistance in integrated devices uncover a significant contribution from the oxygen vacancy defect states induced by IL gating. It provides insights into the evolution of such states with IL gating. In the insulating phase, $S$ behaves as a single-band semiconductor without gating, consistent with band structure calculations [10a]; Under IL gating, however, $S$ deviates strongly from a conventional semiconducting behavior, showing instead a nearly $T$-independent behavior for $V_g$ = 0.5 V and trending towards a near linear temperature dependence at $V_g$ = 1.0 V. Interestingly, charge transport remains thermally activated for all $V_g$ studied, with an activation energy $E_A$ decreasing with increasing $V_g$. The deviation from conventional semiconducting behavior in thermoelectric measurements under IL gating strongly indicates the involvement of oxygen vacancy defect states. The essential behavior of the temperature dependence of Seebeck coefficient and resistance can be reproduced qualitatively by numerical simulation based on a simple model DOS with a defect band residing in the band gap. Our results demonstrate the importance of oxygen vacancy defect states for modulating the MIT in IL-gated $VO_2$, and show that thermoelectric measurement is a sensitive probe for such states.

$VO_2$ thin films with thicknesses of 10~50 nm were deposited on single crystalline sapphire substrates (0001 orientation, MTI Corporation). Film deposition was performed in an electron-beam evaporator (AJA International Inc.) using $VO_2$ pellets (VWR Corporation) as the source. Device geometry was defined using stencil shadow masks. Additional post deposition annealing in oxygen partial pressure in the same vacuum chamber was adopted to compensate for the oxygen deficiency formed during the ebeam deposition. During annealing, $O_2$ and Ar gas flow rate were kept at 6 and 8 sccm, respective, and the total pressure was maintained at 90 mtorr. We found the optimized condition, as characterized by the magnitude of resistance change across the MIT, to be at deposition temperature of 500 °C and post deposition annealing temperature of 550 °C. MIT in samples studied here (15 nm thick) shows a resistance change of up to four orders of magnitude. Device electrodes and heaters for thermoelectric measurements were made by Au deposition using stencil shadow masks. An optical image of a typical device is shown in Fig. 1(a). In order to establish a thermal gradient for the Seebeck coefficient measurement, one side of the sample was placed close to an Au heater, while the other side close to a copper block for anchoring to the base temperature. One droplet (~ 20 ul) of N,N-diethyl-N-(2-methoxyethyl)-N-methylammonium bis-(trifluoromethylsulfonyl)-imide (DEME-TFSI) was dispensed on top of the device to serve as the IL gating media (Fig. 1(a) inset). The ionic liquid was first baked in vacuum at 380 K for over 12 hours to remove trace water in order to minimize the chemical reaction with the sample [16], after which different gate voltages were applied. To ensure reproducibility, we avoided using high gate voltages under which chemical reaction between the sample and IL became significant. Therefore, in this paper we restrict our measurement to $V_g \leq 1.0$ V. In a typical measurement, resistance and Seebeck coefficient were measured simultaneously to ensure that both data are taken under identical gating conditions. Resistances between leads $R_1I+$ and $R_2I+$ were measured using a dc source with alternating polarity. A steady state method was used for Seebeck coefficient measurement. A temperature gradient $\Delta T$ was generated using the patterned Au heater between leads h+ and h-. Two pairs of Au bridges were used for monitoring $\Delta T$ by acting as a resistance thermometer. A gate voltage was applied between leads Vg+ and $R_1$I+, with a picoammeter (Keithley 6485) to monitor the leakage current. Leakage current was found to be less than 3 nA throughout the gating experiments. A negligible contribution from IL to the measured Seebeck coefficient was confirmed by comparing the measurements at $V_g = 0$ for samples with and without IL. The heating current dependence of the thermoelectric voltage $dV$ is shown in Fig. 1(b). A parabolic relation between thermal electric voltage $dV$ vs $I_h$ for both positive and negative heating current ensures an intrinsic thermoelectric response.

Representative temperature dependent resistance and Seebeck coefficient at zero gate voltage ($V_g$ = 0 V) is shown in Fig. 2. A MIT is found at $T_{onset}$ between 325 and 335 K in both $R(T)$ and $S(T)$, with a typical thermal hysteresis reflecting its first-order nature. The transition temperature is close to the typical values in film samples reported earlier [15, 17]. The resistance changes for over three decades across the MIT, showing good sample quality despite its polycrystallinity. In the low-$T$ semiconducting phase, the resistance increases exponentially with decreasing temperature. The temperature dependence is well fitted by $R = R_0 \exp(\frac{E_A}{k_B T})$, as shown by the solid line in Fig. 2(a). This shows that the charge carrier transport in this regime is thermally activated. The corresponding activation energy $E_A$ is found to be 0.19 eV. With band gap of $VO_2$ in the semiconducting phase known to be 0.6 eV [18], this indicates that either a donor level lies at 0.19 eV below the chemical potential for n-type or an accepter level at 0.19 eV above the chemical potential for p-type conduction.

The Seebeck response shows different behaviors above and below the MIT temperature at zero gating. In the metallic phase, $S$ decreases linearly with decreasing $T$, with a value of -40 μV/K at 360 K. Upon cooling below $T_{MIT}$, $S$ increases steeply by more than an order of magnitude to -450 μV/K at 300 K. The values of Seebeck coefficient at both sides of the MIT are in line with earlier reports [19]. The more than ten-fold enhancement in Seebeck response reflects a change of the electronic structure due to the MIT.

For now, we focus on the semiconducting phase, and will return to the interpretation of Seebeck response in the metallic phase later. In a nonmagnetic system, $S$ consists of contributions from both carrier diffusion ($S_d$) and phonon drag ($S_{pd}$). A typical estimate of the temperature below which $S_{pd}$ becomes important is $\theta_D/5$, with $\theta_D$ the Debye temperature. For VO$_2$ $\theta_D$ is around 750K [6], giving $\theta_D/5 \sim 150$K. An earlier thermoelectric study on doped VO$_2$ indeed revealed a phonon drag contribution only below 200 K [20]. Therefore, in the temperature range of our concern (> 200 K), we neglect the phonon drag term and use $S \sim S_d$ as an approximation.

Generally, the linear response Seebeck coefficient is expressed as

$$S = -\frac{k}{e} \int \left(\frac{E-\mu}{k_B T}\right) \frac{\sigma(E)}{\sigma} dE \quad \text{(Eq. 1)},$$

Where $\sigma = \int \sigma(E) dE = e \int g(E) \mu(E) f(E)[1 - f(E)] dE$ is the total conductivity (integrated over all conducting channels), $g(E)$ is the DOS at energy $E$, $\mu(E)$ is the mobility, and $f(E)$ is the Fermi-Dirac distribution function [21]. It is clear that $S$ is inherently sensitive to energy dependence of conductivity. The asymmetry of the conduction above and below $\mu$ determines the sign and magnitude of $S$. In the semiconducting phase, Eq. 1 reduces to

$$S_c = -\frac{k}{e}\left[\frac{E_c - \mu}{k_B T} + A_c\right] \text{ for } E > E_c, \left(A_c = \int_0^\infty \frac{\epsilon}{k_B T} \sigma(\epsilon) d\epsilon / \int_0^\infty \sigma(\epsilon) d\epsilon, \text{ with } \epsilon = E - E_c\right) \quad \text{(Eq. 2)}$$

$$S_v = \frac{k}{e}\left[\frac{\mu - E_v}{k_B T} + A_v\right] \text{ for } E < E_v, \left(A_v = \int_{-\infty}^0 \frac{\epsilon}{k_B T} \sigma(\epsilon) d\epsilon / \int_{-\infty}^0 \sigma(\epsilon) d\epsilon, \text{ with } \epsilon = E_v - E\right) \quad \text{(Eq. 3)}$$

Here, $S_c$ and $S_v$ are the Seebeck response from the conduction band and valence band respectively. The total Seebeck coefficient $S$ is then

$$S = \frac{S_c \sigma_c + S_v \sigma_v}{\sigma_a + \sigma_b}$$

where $\sigma_c$ and $\sigma_v$ are the conductivity contribution from conduction and valence bands, respectively. Note the opposite signs between $S_c$ and $S_v$. For an intrinsic semiconductor with strictly symmetric conduction and valence bands, $S_c$ and $S_v$ would cancel out, leading to zero Seebeck response. For real materials, this is never the case: the asymmetry between the two gives rise to finite thermoelectricity.

We can see that in VO$_2$, $S$ remains negative throughout the temperature and gate voltage ranges. The magnitude of $S$ without gating is also consistent with existing Seebeck coefficient data on VO$_2$ [19, 22]. Negative Seebeck coefficient indicates that the carrier transport is dominated by electrons. Therefore, the carriers with excitation energy of 0.19 eV obtained from the resistance measurements must come from the donor levels below the CBM. Because of the extra carriers excited from the donor levels, Seebeck response of the conduction band $S_c$ dominates. We approximate $S$ as $S_c$, so that Eq. 2 can be used to calculate total $S$. The solid line in Fig. 2(b) shows a good fit by Eq. 2 to the Seebeck coefficient data. The fitting yields $E_C - \mu = 0.09$ eV. Now, a schematic band diagram of VO$_2$ at $V_g = 0$ can be constructed based on the physical parameters obtained from resistance and Seebeck measurements. As shown in the inset of Fig. 2(b), VO$_2$ at zero gating possesses a chemical potential at 0.09 eV below the CBM and 0.19 eV above the donor states, a situation similar to previous reports [19]. The applicability of a single band picture is also consistent with the band structure calculation showing that after the gap opening at $T < T_{MIT}$, $\pi^*$ band is mainly responsible for the electron transport [23].

At finite IL gating voltage, the MIT transition is suppressed to lower temperatures. Meanwhile the resistance change across the MIT reduces by over an order of magnitude for $V_g = 1.0$ V (shown in Fig. 3). The conduction remains to be thermally activated, as shown by the solid lines in Fig. 3(a). However, the activation energy $E_A$ decreases with increasing IL gating voltage (Fig. 3(a) inset). The reduced $E_A$ suggests a shift of the donor levels towards the chemical potential due to IL gating.

The thermoelectric response at finite IL gating in Fig. 3(b) delivers rich insight. Let's first focus on the Seebeck coefficient in the metallic phase. As can be seen in Fig. 4(a), $S$ is a linear function of $T$ for all gating conditions in our experiments. In a metallic state, the general expression of Seebeck coefficient eq. 1 is reduced to the well-known Mott formula [21]

$$S = -\frac{\pi^2}{3}\frac{k_B}{e}k_B T\left[\frac{d\ln(\mu(E)g(E))}{dE}\right]_{E=E_F} = -\frac{\pi^2}{3}\frac{k_B}{e}k_B T\left[\frac{d\ln(\sigma(E))}{dE}\right]_{E=E_F} \quad \text{(Eq. 4),}$$

where $k_B$ is the Boltzmann constant, $\sigma(E)$ the energy dependent conductivity and $E_F$ the Fermi energy. It is worth noting that the slope decreases as $V_g$ increases from 0 to 0.5V, but remains nearly constant as $V_g$ is further increased to 1.0 V. By assuming a degenerate carrier population of the band, Eq. (4) can be simplified as [24], $S = \frac{\pi^2 k^2 T}{3eE_F}\left(\frac{3}{2} + r\right)$, where $r$ is the power-law index of the carrier scattering time. Take $r = -1/2$ (the phonon scattering dominated case), the slope in the fitting in the inset of Fig. 3(b) corresponds to $E_F \sim 0.11$eV for $V_g = 0$, and ~0.17 eV for $V_g = $ 0.5 and 1.0 V, respectively. The increase of $E_F$ with increasing $V_g$ reflects the increase in carrier density upon gating. We can further estimate the carrier density using $E_F = \frac{h^2}{8m^*}\left(\frac{3n}{\pi}\right)^{2/3}$, with the effective carrier mass $m^* = 3m_0$ as reported earlier [6], carrier density is found to be ~1.82×10$^{22}$

cm$^{-3}$ for $V_g = 0$. The estimated carrier density at zero gating is very close to the value reported previously using Hall measurement [14].

In the semiconducting state, the Seebeck response under IL gating changes dramatically from that without gating. As shown in Fig. 3 (b), at $V_g = 0.5$ V, $S(T)$ shows a nearly $T$-independent behavior. For higher $V_g = 1.0$ V, $S$ changes slope into decreasing with decreasing $T$, resembling a metallic temperature dependence. The corresponding Peltier coefficient $\Pi = S \cdot T$ is shown in Fig. 3(c). The temperature independent part of $\Pi$ quantifies the energy gap of $-(E_c - \mu)/e$ (for electrons), which is roughly 0.09 V for ungated VO$_2$. $\Pi$ reduces to half of its value at $V_g = 0.5$ V, and becomes very small at $V_g = 1.0$ V. The rapid decrease of Peltier coefficient upon gating indicates the increase of the chemical potential of VO$_2$ in the semiconducting phase.

The temperature dependence of conductivity and Seebeck coefficient in the semiconducting phase show distinctively different behavior upon IL gating, which is counter intuitive at first glance. This difference can be understood as following: [25] it is the electrons right at the chemical potential that contribute the most to conductivity; while for thermoelectric transport, it is the asymmetry of electron occupation below and above the chemical potential that matters. Therefore, the contrasting behaviors between Seebeck coefficient and resistivity signify a change in the DOS near the chemical potential in VO$_2$ under IL gating.

At $V_g = 0$, the donor defect level, due to oxygen vacancies in the as-grown VO$_2$ and perhaps other defects, has very narrow dispersion. Its role is to provide thermally excited carriers. The donor defect states do not participate in transport directly due to the fact that carriers in these states are highly localized. Since the conduction band dominates the carrier transport, $S(T)$ can be described by Eq. (2) with a negative slope. By increasing $V_g$ to 1 V, the opposite slope of $S(T)$ requires participation of states below the chemical potential $\mu$, whose contribution overtakes that from the conduction band. As the valence band is far away from $\mu$ and its contribution to $S$ is not expected to change much with $V_g$, it is very likely that the additional DOS originates from the broadening of the donor level into a defect band, where carriers become mobile and contribute directly to transport.

As investigated in IL gating experiments under controlled atmosphere [15], creation of oxygen vacancies (Vo) are found to be the major effect in addition to the electrostatic gating. Our measurements show that the chemical potential shifts closer to the conduction band edge with increasing $V_g$. This could come from two factors: 1) The defect level is broadened due to the higher concentration of Vo created at higher $V_g$. 2) The bulk CBM is distorted by the Vo defects, leading to tail states extending well into the band gap. At $T = 0$, the defect states are fully occupied. However, at finite temperatures the carriers in the defect states can be ionized to push the chemical potential $\mu$ closer to the now distorted CBM, and even into the tail states of the conduction band, since these tail states are expected to have significantly lower DOS than that of the bulk CBM.

With the above physical picture, the essence of the contrasting thermoelectric response and carrier transport can be captured by a numerical simulation using Eq. 1 for Seebeck coefficient (and conductivity formula therein). We use here a simple model DOS, in which the valence bands are ignored. The DOS of defect states and the bulk DOS (with the distorted CBM tail) both are modeled with a Gaussian function $a\sqrt{\frac{c}{\pi}}e^{-c(x-b)^2}$. The relative magnitude of the integration of the two reflects the IL gating dependent defect density, characterized by the coefficient $a$. Parameters $b$ and $c$ set the position and width of the state, respectively. Assuming that the mobility is weakly dependent on energy near the Fermi level, the mobility term in Eq. 1 is treated as a constant for both the defect and conduction bands. A representative DOS we used for the calculation is shown in Fig. 4(a). Based on this DOS, we carried out numerical integration to find the chemical potential as a function of temperature, which is then used in Eq. 1 for further calculations of $R(T)$ and $S(T)$. Fig. 4(b) shows the calculated $T$-dependent resistance and Seebeck coefficient. It captures the key experimental observation: a thermally activated $R(T)$ and an apparent 'metal-like' $S(T)$. Thus, it is clear that the nature of the contrasting thermoelectric response and carrier transport in $VO_2$ under IL gating is purely a band structure property.

Combining thermoelectric and carrier transport probes, which detect different parts of DOS near the chemical potential, our work leads to correct attribution of suppressed MIT in $VO_2$ to IL gating induced defect states. It would otherwise evade the detection in a typical transport measurements [14], until detailed oxygen isotope experiment further pointed out the additional defect states upon IL gating[15]. Our approach thus provides a straightforward and effective method that can differentiate the contribution from the electrostatic gating and those from the induced defect states.

In conclusion, simultaneous charge and thermoelectric transport measurements were performed in IL gated VO$_2$ thin films. We obtained clear signature of IL gating induced defect states, due to the formation of oxygen vacancies. The ability of differentiating the contribution from the electrostatic gating and those from the induced defect states make our measurements an effective approach in probing the intrinsic transport mechanisms, which may find more applications in a wide range of materials and field effect transistor devices involving ionic liquid.

The authors acknowledge Weidong Luo for useful discussions. The work done in China is supported by National Key Projects for Research & Development of China (Grant No. 2019YFA0308602), National Natural Science Foundation of China (Grant No. 11804220). Work at SUNY-Buffalo is supported by NSF under Grant No. DMR-1229208. H.X. also acknowledges additional support from a Shanghai talent program.

Figure captions:

Fig.1 (a) An optical image of a typical device on sapphire substrate. The light brown colored stripe in the center is the 15 nm thick $VO_2$ film. Inset shows the ionic droplet on top of the device. (b) A representative heating power dependence of the Seebeck voltage at 300K. The Seebeck voltage scales linearly with heating power.

Fig. 2 (a) The temperature dependence of resistance $R(T)$ at $V_g = 0$. The solid line is the fitting of a thermally activated temperature dependence $R = R_0 \exp(\frac{E_A}{k_B T})$. Open and solid symbols stand for the cooling and warming curves, respectively. (b) The temperature dependence of Seebeck coefficient $S(T)$ at $V_g = 0$. The solid line is a fitting by Eq. 2 in the main text. The inset shows a band diagram of $VO_2$ in the semiconducting phase with parameters extracted from the measurements.

Fig. 3 (a) The temperature dependence of resistance $R(T)$ at different gate voltages. Solid lines are the fitting with a thermally activated temperature dependence. The corresponding energy gap are shown in the inset. Open and solid symbols stand for the cooling and warming curves, respectively. (b) The temperature dependence of Seebeck coefficient $S(T)$ at different gate voltages. Inset shows the high-$T$ part of $S(T)$, with the linear fitting using the Mott formula (see main text). (c) The temperature dependence of Peltier coefficient $\Pi(T)$ at different gate voltages.

Fig. 4 (a) The modeled density of states used in the calculation. The small peak around $E = 0$ (the defect state) and the conduction band are modeled by a Gaussian function $a\sqrt{\frac{c}{\pi}}e^{-c(x-b)^2}$, with parameters (a, b, c) of (0.3, 0, 0.008) and (5, 0.2, 0.1) respectively. (b) The calculated temperature dependence of resistance and Seebeck coefficient.

Figures:

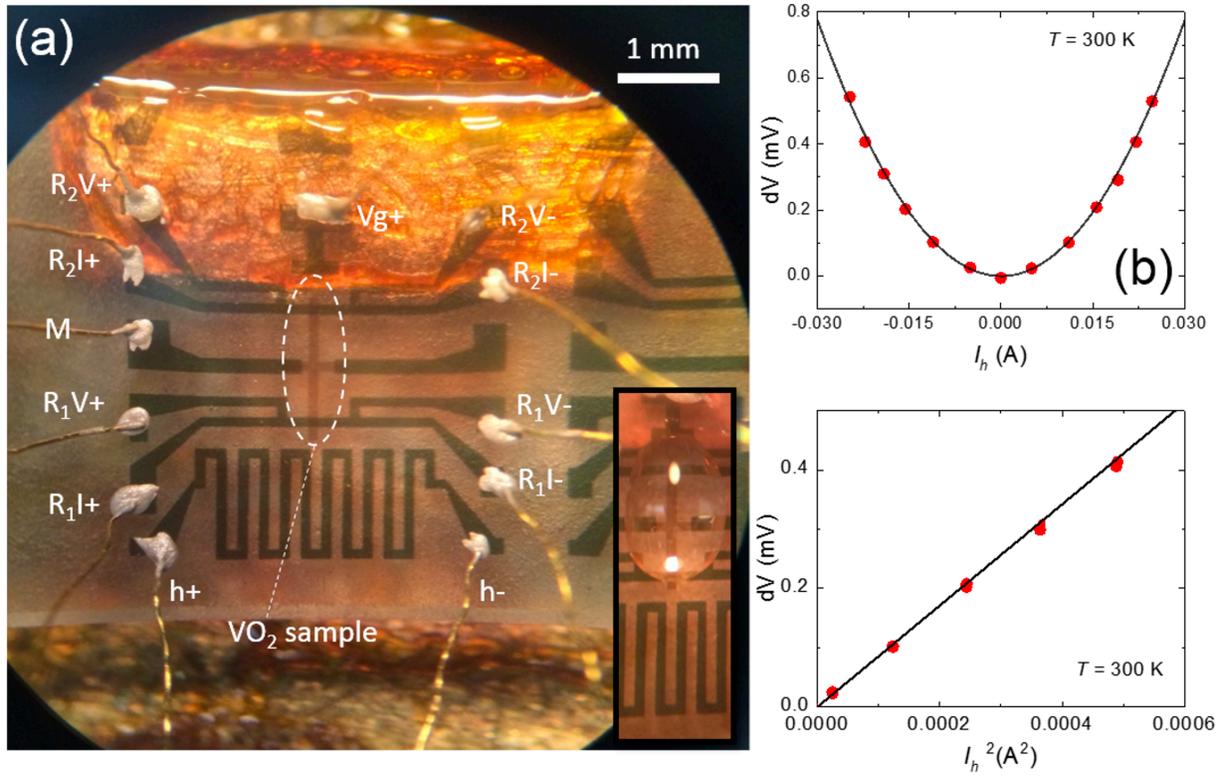

Fig. 1

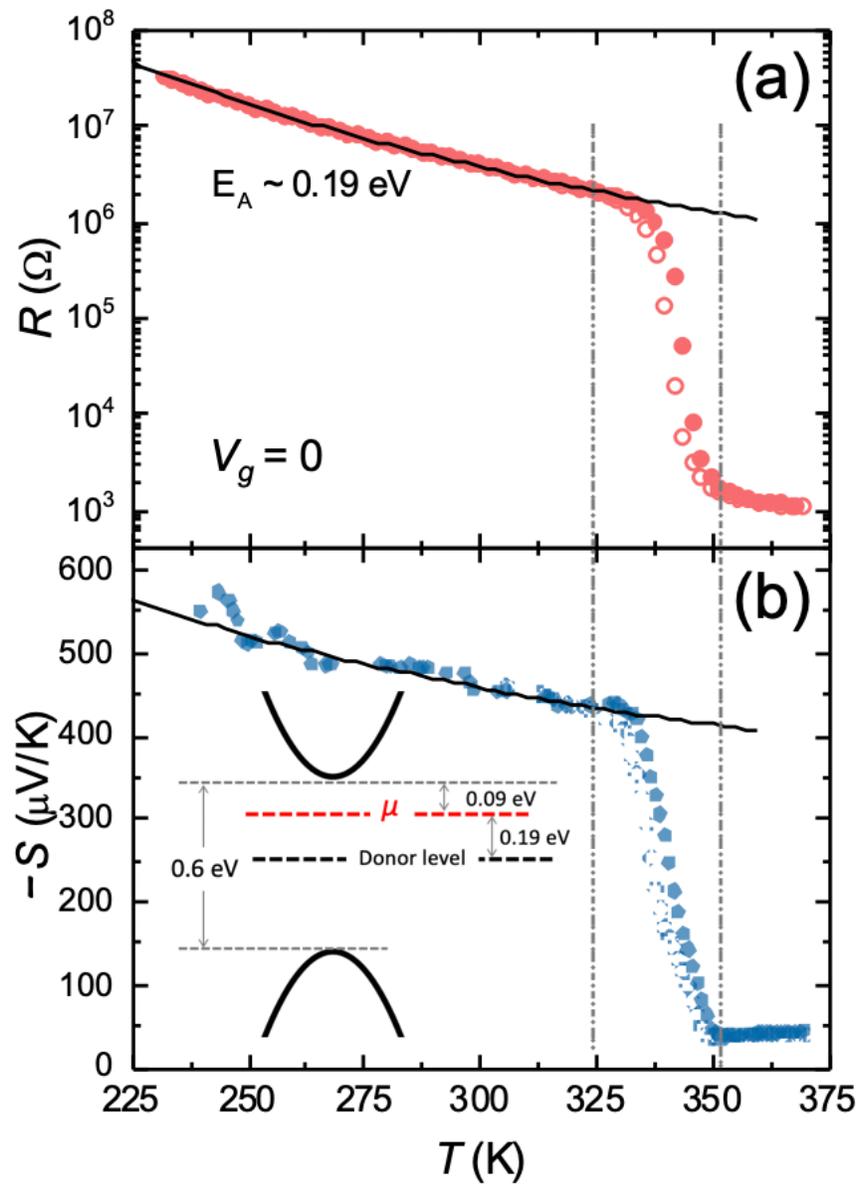

Fig. 2

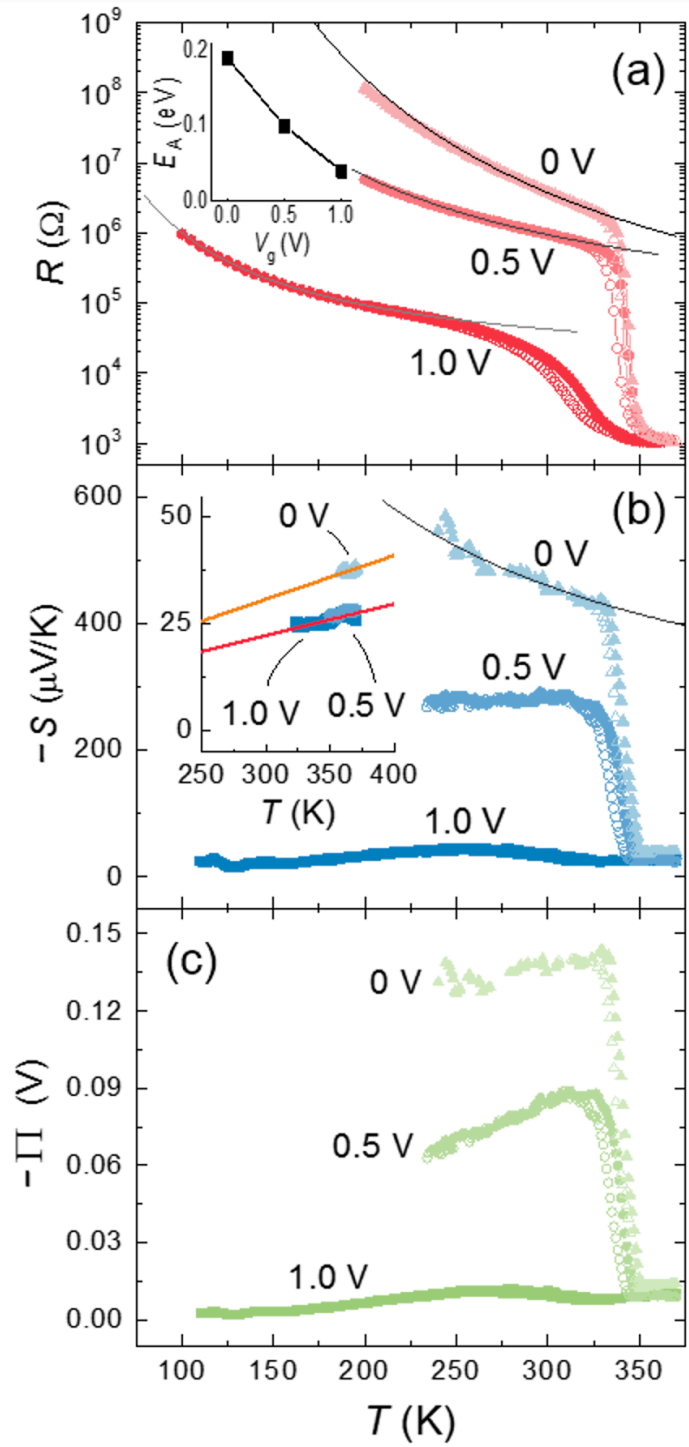

Fig. 3

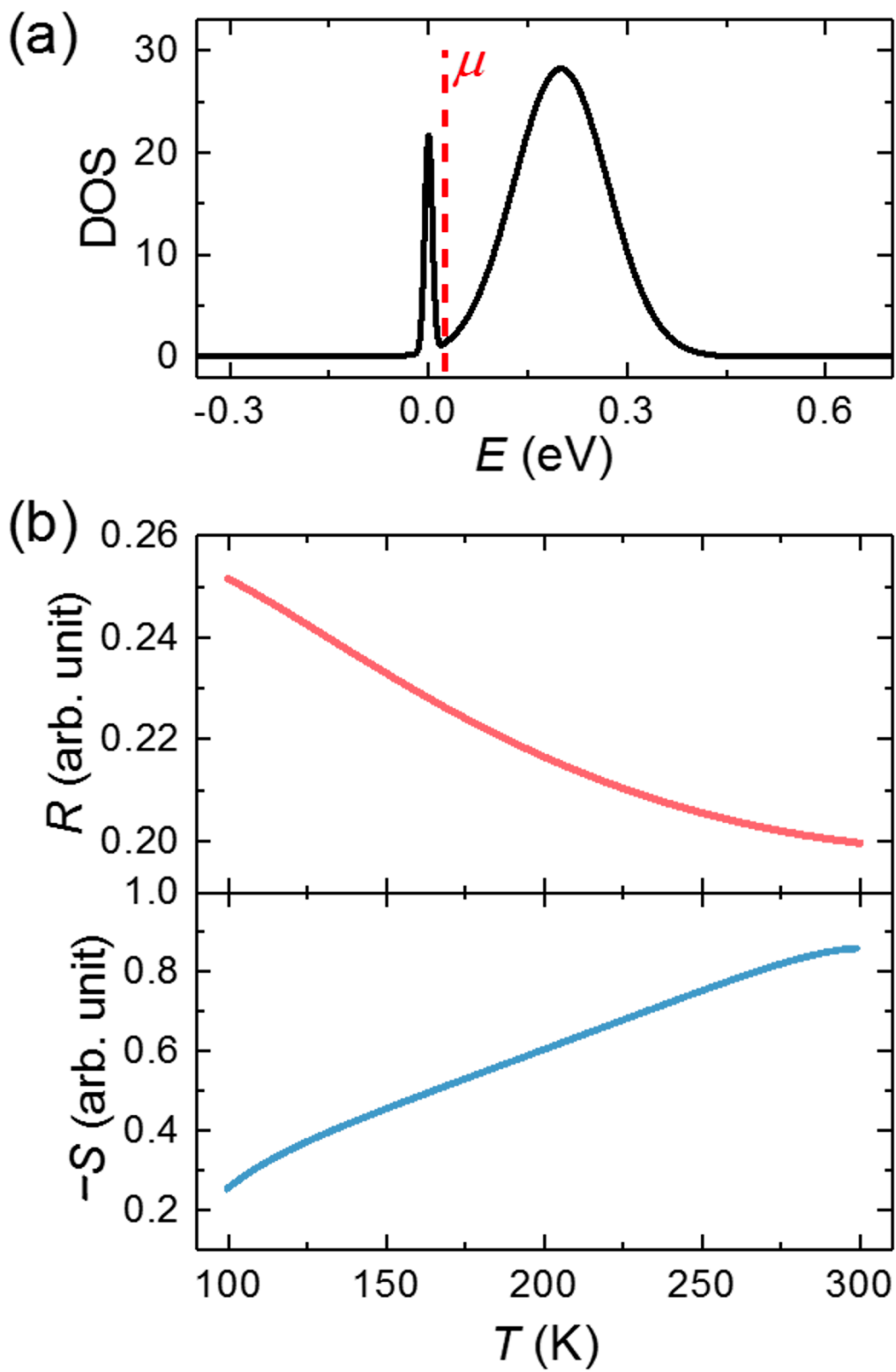

Fig. 4

# References:

[1] F. J. Morin, *Physical Review Letters* **1959**, 3, 34.
[2] R. M. Wentzcovitch, W. W. Schulz, P. B. Allen, *Physical Review Letters* **1994**, 72, 3389.
[3] a) B. Hu, Y. Ding, W. Chen, D. Kulkarni, Y. Shen, V. V. Tsukruk, Z. L. Wang, *Advanced Materials* **2010**, 22, 5134; b) S. M. Babulanam, T. S. Eriksson, G. A. Niklasson, C. G. Granqvist, *Solar Energy Materials* **1987**, 16, 347; c) N. B. Aetukuri, A. X. Gray, M. Drouard, M. Cossale, L. Gao, A. H. Reid, R. Kukreja, H. Ohldag, C. A. Jenkins, E. Arenholz, K. P. Roche, H. A. Durr, M. G. Samant, S. S. P. Parkin, *Nat Phys* **2013**, 9, 661.
[4] a) S. Biermann, A. Poteryaev, A. I. Lichtenstein, A. Georges, *Physical Review Letters* **2005**, 94; b) M. W. Haverkort, Z. Hu, A. Tanaka, W. Reichelt, S. V. Streltsov, M. A. Korotin, V. I. Anisimov, H. H. Hsieh, H. J. Lin, C. T. Chen, D. I. Khomskii, L. H. Tjeng, *Physical Review Letters* **2005**, 95, 196404.
[5] V. Eyert, *Annalen der Physik* **2002**, 11, 650.
[6] C. N. Berglund, H. J. Guggenheim, *Physical Review* **1969**, 185, 1022.
[7] a) S. Wall, S. Yang, L. Vidas, M. Chollet, J. M. Glownia, M. Kozina, T. Katayama, T. Henighan, M. Jiang, T. A. Miller, D. A. Reis, L. A. Boatner, O. Delaire, M. Trigo, *Science* **2018**, 362, 572; b) V. R. Morrison, R. P. Chatelain, K. L. Tiwari, A. Hendaoui, A. Bruhacs, M. Chaker, B. J. Siwick, *Science* **2014**, 346, 445; c) B. T. O'Callahan, A. C. Jones, J. H. Park, D. H. Cobden, J. M. Atkin, M. B. Raschke, *Nat Commun* **2015**, 6, 6849.
[8] S. Lee, K. Hippalgaonkar, F. Yang, J. W. Hong, C. Ko, J. Suh, K. Liu, K. Wang, J. J. Urban, X. Zhang, C. Dames, S. A. Hartnoll, O. Delaire, J. Q. Wu, *Science* **2017**, 355, 371.
[9] L. Chen, Z. J. Xiang, C. Tinsman, T. Asaba, Q. Huang, H. D. Zhou, L. Li, *Applied Physics Letters* **2018**, 113, 061902.
[10] a) J. Karel, C. E. ViolBarbosa, J. Kiss, J. Jeong, N. Aetukuri, M. G. Samant, X. Kozina, E. Ikenaga, G. H. Fecher, C. Felser, S. S. P. Parkin, *ACS Nano* **2014**, 8, 5784; b) Y. Yamada, K. Ueno, T. Fukumura, H. T. Yuan, H. Shimotani, Y. Iwasa, L. Gu, S. Tsukimoto, Y. Ikuhara, M. Kawasaki, *Science* **2011**, 332, 1065; c) J. Wei, H. Ji, W. H. Guo, A. H. Nevidomskyy, D. Natelson, *Nat Nanotechnol* **2012**, 7, 357; d) J. D. Budai, J. Hong, M. E. Manley, E. D. Specht, C. W. Li, J. Z. Tischler, D. L. Abernathy, A. H. Said, B. M. Leu, L. A. Boatner, R. J. McQueeney, O. Delaire, *Nature* **2014**, 515, 535.
[11] K. Ueno, S. Nakamura, H. Shimotani, A. Ohtomo, N. Kimura, T. Nojima, H. Aoki, Y. Iwasa, M. Kawasaki, *Nat Mater* **2008**, 7, 855.
[12] B. Skinner, M. S. Loth, B. I. Shklovskii, *Physical Review Letters* **2010**, 104, 128302.
[13] H. Yuan, H. Shimotani, A. Tsukazaki, A. Ohtomo, M. Kawasaki, Y. Iwasa, *Advanced Functional Materials* **2009**, 19, 1046.
[14] M. Nakano, K. Shibuya, D. Okuyama, T. Hatano, S. Ono, M. Kawasaki, Y. Iwasa, Y. Tokura, *Nature* **2012**, 487, 459.
[15] J. Jeong, N. Aetukuri, T. Graf, T. D. Schladt, M. G. Samant, S. S. P. Parkin, *Science* **2013**, 339, 1402.
[16] H. Ji, J. Wei, D. Natelson, *Nano Letters* **2012**, 12, 2988.
[17] J. H. Park, J. M. Coy, T. S. Kasirga, C. Huang, Z. Fei, S. Hunter, D. H. Cobden, *Nature* **2013**, 500, 431.
[18] H. W. Verleur, A. S. Barker, C. N. Berglund, *Physical Review* **1968**, 172, 788.
[19] a) J. Cao, W. Fan, H. Zheng, J. Wu, *Nano Letters* **2009**, 9, 4001; b) D. Fu, K. Liu, T. Tao, K. Lo, C. Cheng, B. Liu, R. Zhang, H. A. Bechtel, J. Wu, *Journal of Applied Physics* **2013**, 113, 043707.
[20] J. M. Reyes, M. Sayer, A. Mansingh, R. Chen, *Canadian Journal of Physics* **1976**, 54, 413.
[21] H. Fritzsche, *Solid State Communications* **1971**, 9, 1813.
[22] T. Katase, K. Endo, H. Ohta, *Physical Review B* **2015**, 92, 035302.
[23] J. B. Goodenough, *Journal of Solid State Chemistry* **1971**, 3, 490.

[24]　D. K. C. MacDonald, *Thermoelectricity: an introduction to the principles*, John Wiley & Sons, Inc., New York **1962**.
[25]　J. M. Ziman, *Principles of the theory of solids*, Cambridge University Press, **1972**.